\documentclass[12pt]{article}
\usepackage{graphicx,amsmath}
\usepackage{color}

\newlength{\dinwidth}
\newlength{\dinmargin}
\def\be{\begin{equation}}   
\def\ee{\end{equation}}  
\def\bea{\begin{eqnarray}}                      
\def\eea{\end{eqnarray}}
\def\ch1{$\chi(1^+)$}
\def\lapproxeq{\lower .7ex\hbox{$\;\stackrel{\textstyle                                                    
<}{\sim}\;$}}                                                    
\def\gapproxeq{\lower .7ex\hbox{$\;\stackrel{\textstyle                                                    
>}{\sim}\;$}}

\begin{document}

\begin{flushright}                                                    
IPPP/15/40  \\
DCPT/15/80 \\                                                    
\today \\                                                    
\end{flushright} 

\vspace*{0.5cm}

\begin{center}
{\Large \bf Inclusive prompt $\chi_{c,b}(1^{++})$ production at the LHC}\\
\vspace{1cm}

A.G. Shuvaev$^a$, V.A. Khoze$^{a,b}$, A.D. Martin$^b$ and  M.G. Ryskin$^{a,b}$.\\ 

\vspace{0.5cm}

$^a$ Petersburg Nuclear Physics Institute, NRC Kurchatov Institute, Gatchina, St.~Petersburg, 188300, Russia \\ 
$^b$ Institute for Particle Physics Phenomenology, University of Durham, Durham, DH1 3LE \\ 

\begin{abstract}
We study the prompt production of the $\chi_c(1^+)$ and $\chi_b(1^+)$ mesons at high energies. Unlike $\chi(0^+,2^+)$ production, $\chi (1^+)$ mesons cannot be created at LO via the fusion of two on-mass-shell gluons, that is $gg\to\chi_{c,b}(1^+)$ are not allowed. However, the available experimental data show that the cross sections for $\chi_c(1^+)$ and $\chi_c(2^+)$ are comparable. We therefore investigate four other $\chi(1^+)$ production mechanisms: namely, (i) the standard NLO process $gg\to\chi_{c,b}(1^+)+g$, (ii) via gluon virtuality, (iii) via gluon reggeization and, finally, (iv) the possibility to form $\chi_{c,b}(1^+)$
by the fusion of three gluons, where one extra gluon comes from another
parton cascade, as in the Double Parton Scattering processes.
\end{abstract}

\end{center}
\vspace{0.5cm}

\section{Introduction}
It is well known that according to the generalised Landau--Yang selection rule a spin-1 meson cannot be produced in the fusion of two  identical massless spin-1 particles~\cite{BLP}. Therefore the inclusive $\chi_c(1^+)$ cross section 
 offers the possibility to probe some non-trivial dynamics of the 
NLO interaction.

There are two ways to overcome the Landau-Yang selection rule: either to account for different virtualities of incoming gluons (that is to violate their identity) or to consider the formation of  $\chi(1^+)$ by three gluons. The third gluon may be emitted as the new secondary particle $gg\to \chi_(1^+)+g$ process, or it may occur in the initial state in the fusion process $g+(gg)\to\chi$ fusion.
Recall that, as shown in~\cite{jpsi1}, the role of such three gluon mechanisms of $J/\psi$-meson production increases with energy\footnote{Unfortunately there was a confusion in \cite{jpsi1} -- for the 3-gluon mechanism the cross section, and not the amplitude, grows as ln$s$. Thus the cross section at LHC energies will be a few times smaller.} and may even be dominant at the very high energies. Recently the three gluon  $J/\psi$ production was also studied in~\cite{mot}. Note that, in comparison with the $J/\psi$,
 the dynamics of $\chi_c(1^+)$ formation is richer since,
 due to its negative $C$-parity, the $J/\psi$ meson cannot be produced 
in the fusion of {\em two} gluons, even if we account for the different virtualities of these gluons.

The available  experimental data on prompt production of $\chi_c$
mesons in high energy hadronic collisions
(see \cite{chi-data} for a recent review)
concern, as a rule, the transverse momentum $p_t$ distribution of
the $\chi$-meson production rates or the ratios
of $\sigma(\chi_{c,b}(2^+ ))/\sigma(\chi_{c,b}(1^+ ))$ as a function of $p_t$
\footnote{Experimentally it is easier to measure 
such ratios since various theoretical
and experimental uncertainties cancel out.
As far as we aware, currently there are no high-energy data at low $p_t$ and no data on the
rapidity distribution  $d\sigma/dy$ or on the total cross sections
for $\chi_{c,b}(1)$ except for the relatively low energy data in $\pi Be$ collisions.
 Note that  the $p_t$ of the $J/\psi$ and not
that of $\chi_c$ is measured, but due to the large $J/\psi$ mass
(close to the $\chi_c$ mass) the $p_t$ of $J/\psi$ is rather close to that of $\chi_c$, and qualitatively 
reproduces well the behaviour with respect to the $p_t$ of $\chi_c$}.
 The ratio $R_{21}=\chi(2^+)/\chi(1^+ )$ is of the order of one and practically does not depend on $p_t$. It can be
 described by a constant, $R_{21}$, independent of $p_t$, both for $\chi_c$ and $\chi_b$ 
processes: in particular for $\chi_b$ the value $R_{21}=0.85\pm 0.07$~\cite{chib}.\footnote {Note that there is an indication
for a  rise of the ratio $R_{21}$ at low $p_t$
in the LHCb measurement of $\chi_c$ at 7 TeV \cite{Aaij:2013dja}. Therefore it is desirable that the LHCb collaboration
perform a new  measurement of this ratio at low $p_t$
during run-II of the LHC.}
This was  not expected. The lowest-order theory predicts the growth of this ratio for decreasing $p_t$,
  since, at low $p_t$, $\chi(2)$ meson can be produced at  LO
 while $\chi(1)$ only occurs at NLO (see, for example, ~\cite{LLP}).
Moreover, in fixed-target $\pi-Be$ interactions at $515$ GeV/c  the $p_t$-integrated yields
of $\chi_c(1^+)$ ($\sigma=232\pm 37\pm 37$ nb) and $\chi_c(2^+)$ ($\sigma=407\pm 71\pm 69$) 
nb~\cite{Be} were measured. This does not indicate the
 expected  {\em strong} suppression of $\chi_c(1)$ 
which cannot be produced at LO via on-mass-shell gluon-gluon fusion. 
 It is worth mentioning that new important information
could come from fixed-target experiments using an LHC
 beam, as advocated
 in \cite{Massacrier:2015qba}, where lower $p_t$ values
could be reached.
Thus it is topical to study theoretically the different possible mechanisms of 
$\chi(1)$ production. 

The inclusive cross sections were calculated at lowest order  in~\cite{GTW}. 
Numerical evaluations at the LHC energy are given, for example, in~\cite{LLP}.
 	A specific higher-order process - so-called `$s$-channel $Q\bar Q$-cut'
was considered in \cite{lans1,lans2,qq-cut}. For a more detailed review see, for example,~\cite{rev}.

In the present paper we discuss all the different possibilities of colour-singlet $\chi_c(1^+)$ and $\chi_b(1^+)$ inclusive production. We will not discuss here the colour-octet contribution, in particular, since 
at the moment the importance of colour-octet
contributions to the $\chi_c(1,2)$ production remains open, see
for instance~\cite{chi-data,rev,  Shao,octet}.
Note that Ref.~\cite{Shao} presents
the complete NLO NRQCD predictions for the polarized
 $\chi_c(1,2)$ production at medium and high $p_t$, which
when compared to forthcoming LHC measurements could allow a detailed
probe of the validity of NRQCD and colour-octet  mechanism.

   We consider the lowest $\alpha_s$ order
process and, moreover,  if in some kinematical domain we find that the cross section is 
enhanced by a large logarithm (either of virtuality, $\ln q^2$, or of energy, $\ln s$), then we will focus on this leading logarithm (LL) contribution. We emphasize that the aim of the paper is not to present a precise  quantitative prediction\footnote{Precise results would require dedicated higher-order calculations.}, but rather to compare the role of different mechanisms of \ch1 production in high energy hadron-hadron collisions.

 We note that $\chi_b$ production has advantages over $\chi_c$  both from the theoretical and experimental viewpoints. Due to the larger mass of $\chi_b$ (i) the perturbative QCD approach is better justified and (ii) it is easier to detect the more energetic decay products.

In Sect.~2 we first consider the  cross section caused by the $gg\to\chi(1^+)+g$ subprocess, and secondly due to the different virtualities of two incoming gluons. Then in Sect.~3 we discuss the production via  three-gluon fusion where a pair of $t$-channel gluons
represents the reggeization of one incoming gluon. Next, in Sect.~4 we study a most interesting possibility when the two gluons come from two different parton cascades. This mechanism for \ch1 production has not been studied before.  At asymptotically high energy this should be the dominating contribution. An interesting fact is that in such a case the cross section in the central region is expected to be smaller than that near the proton fragmentation domain. We give our conclusions  in Sect.~5.

\section{Lowest order $\chi(1^+)$ production}
\begin{figure}
\begin{center}
\includegraphics[height=6cm]{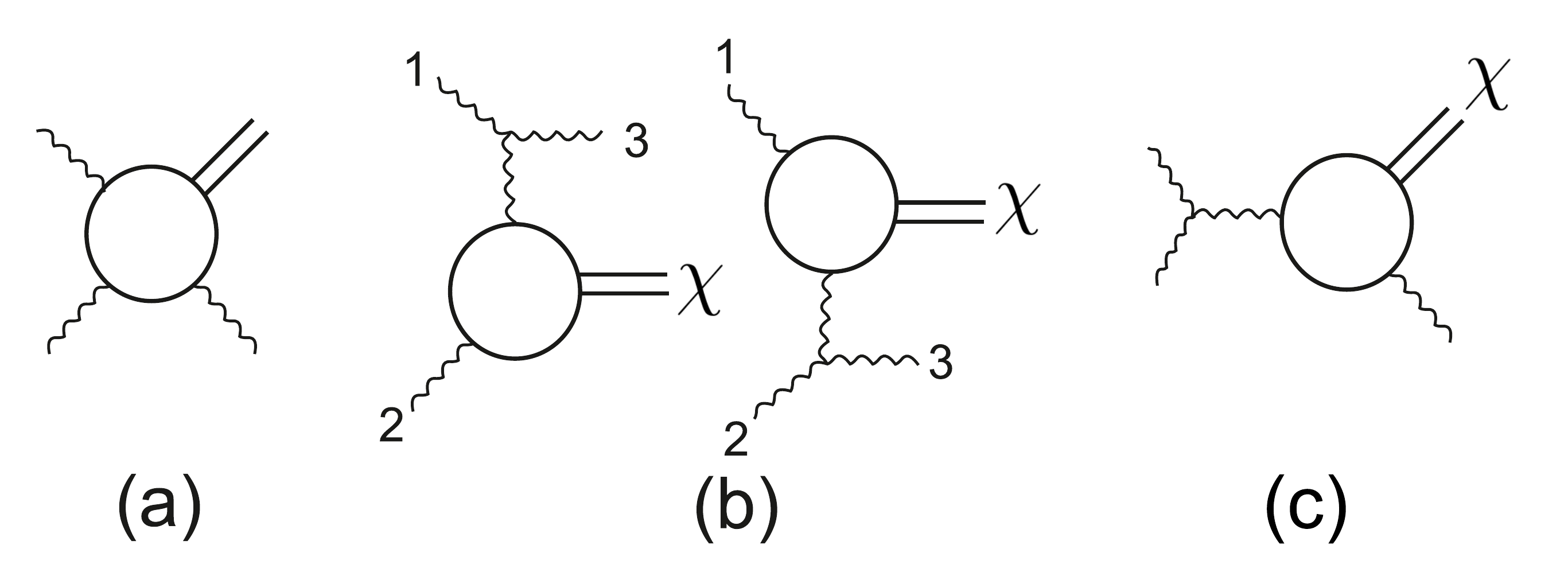}
\caption{\sf Subprocesses for $\chi_(1^+)$ production: (a) the standard $gg\to  \chi(1^+) + g$ process where three on-mass-shell gluons couple to the heavy quark loop; (b,c) where one virtual gluon and one on-mass-shell gluon couple to the heavy quark loop --- the virtual gluon has $q^2<0$ for (b) and $q^2>0$ for (c).
}
\label{fig:proc12}
\end{center}
\end{figure}

In this section we study production by the two-gluon initiated states shown by `1 and 2' 
 in Fig.~\ref{fig:proc12}. We show separately the contribution where all three gluons couple to 
the heavy quark loop (Fig.1a), and the contribution where only two $t$-channel gluons couple to the heavy quark loop (Fig.1b) which may not vanish in for $\chi(1^+)$ if the virtualities of these two
 gluons are different (Fig.1b).
 This contribution may be considered separately and  formally called `production' via two
 gluon fusion". However actually this is just part of the whole $gg\to \chi(1^+)+g$ cross section. 
 
 Another point is that it is natural to consider for this part the role of $t$-channel gluon reggeization, which is the simplest example of \ch1 production via 3-gluon fusion $g+(gg)\to \chi(1^+)$. For this reason we will describe the contribution of Fig.~1b in more detail in Section~\ref{sec:reg}, and the reggeization in Section~\ref{sec:Regge}.

 Finally, in Fig.1c we show the process initiated by a highly virtual $s$-channel gluon.

\subsection{The $gg\to  \chi(1^+) + g$ process}
The simplest possibility to overcome the Landau-Yang selection rule is to create the \ch1 meson together with an additional gluon. The corresponding `hard' cross section was calculated in~\cite{GTW}
$$\frac{d\hat\sigma(gg\to\chi(1^+)+g)}{dt}=\frac{12\pi\alpha^3_s R^{'2}}{M^3s^2}\cdot  $$
\be
\cdot\frac{P^2[M^2P^2(M^4-4P)+2Q(-M^8+5M^4P+P^2)-15M^2Q^2]}{(Q-M^2P)^4}\ ,
\label{eq:eq1}
\ee
where $s,t,u$ are the Mandelstam variables for the hard subprocess. The quantities $P=st+su+ut$ and $Q=stu$; and $M$ is the $\chi_c(1^+)$ or $\chi_b(1^+)$ meson mass.

In eq.(\ref{eq:eq1}) the nonrelativistic wave function of the meson is assumed.  The derivative of the $P$-wave wave function at origin is denoted as $R'$. In our calculations for $\chi_c(1^+)$ we use the values $R^{'2}/M^2=0.006$ GeV$^3$ (as in \cite{GTW}) and the QCD coupling $\alpha_s=0.335$ which provide reasonable widths of the $\chi_c(0,1,2)$ mesons calculated accounting for the $\alpha_s$ corrections (see e.g.~\cite{Dia}). For $\chi_b(1^+)$ we take $R^{'2}=1$ GeV$^5$ similar to that in \cite{LLP}, which is consistent with the potential model results, and $\alpha_s=0.22$ corresponding to the higher scale appropriate for $\chi_b$ production.

In comparison with the natural parity $\chi_c(0^+)$ and $\chi_c(2^+)$ mesons, 
which can be formed via on-mass-shell gluon-gluon fusion with the cross section $\hat\sigma\propto \alpha^2_s$, the cross section (\ref{eq:eq1}) contains and extra power of QCD coupling (not accompanied by any large logarithm). Therefore $gg\to  \chi(1^+) + g$ can be considered as a NLO production process.

Expression (\ref{eq:eq1}) has to be convoluted with the incoming parton distributions and integrated over $t$ and $s$. Since we are looking for the inclusive \ch1 production we fix the rapidity, $Y$, of \ch1 meson and choose to integrate over the variables $y$ and $p_t$, where $p_t$ is the transverse momentum of the final gluon while $y$ is the rapidity separation between this gluon and the \ch1 meson. It is easy to check that the corresponding Jacobian $J=1$ (see e.g. \cite{KMR-P}). Thus the cross section of \ch1 production in proton-proton collisions reads
\be
\frac{d\sigma}{dY}=\int x_1g(x_1,\mu_F)\frac{d\hat\sigma(gg\to\chi(1^+)+g)}{dt}x_2g(x_2,\mu_F) dy dp_t^2\ ,
\label{eq:eq2}
\ee
where $g(x,\mu_F)$ is the density of gluons which carry a fraction $x$ of the momentum of the incoming proton (measured at factorization scale $\mu_F$). The variables 
\be
x_{1,2}=(m_t+p_t e^{\pm y})e^{\pm Y}/\sqrt s
\ee
where $m_t^2=M^2+p^2_t$. The hard cross section $d\hat\sigma/dt$ is given by (\ref{eq:eq1}) with 
\bea
s & = & m^2_t+p^2_t+m_tp_t(e^y+e^{-y})  \\
t & = &-p^2_t-m_tp_te^y \\
u& = & -p^2_t-m_tp_te^{-y}.
\eea
The expected cross section is shown in Table 1 
  where the LO MSTW2008~\cite{MSTW08}  PDFs were used.

\subsection{Production via $g^*+g\to \chi(1^+)$ fusion  \label{sec:reg}}
Here we consider the possibility of \ch1 production via `virtual$+$real' gluon fusion processes\footnote{The production via  $g^*+g$ fusion was discussed in detail in \cite {Hagler:2000dd}.}, which are shown in Fig.~\ref{fig:proc12}. These processes are not forbidden thanks to the different
 mass/virtualities of the initial gluons. The corresponding `hard' 
cross section can be extracted from (\ref{eq:eq1}) using the `equivalent photon/gluon' approximation~\cite{WW}. Indeed, in the limit of $s\gg M^2,|t|$, the dominant contribution comes from a diagram where first the incoming gluon with momentum, say, $p_1$ emits the final gluon $p_3$ and then the virtual gluon, $g^*$, with momentum $q=p_1-p_3$ interacts with another incoming (quasi-real) gluon $p_2$ to produce the \ch1 meson~\footnote{
Recall that in terms of the unintegrated gluon density $f_g(x,q_t,\mu_F)$
the value of $xg(x,\mu_F)$ is given by the logarithmic integral
$
xg(x,\mu_F)=\int^{\mu_F} f_g(x,q_t,\mu_F)~dq^2_t/q^2_t.
$
That is, each function $xg$ contains a large logarithm. In this logarithmic integration the virtuality of the initial gluon is small, $q^2\sim q^2_t \ll \mu^2_F$. So this gluon may be considered as an on-mass-shell particle.}.

Thus in the large $s$ limit we may write the cross section (\ref{eq:eq1}) as the product of the virtual gluon flux, $dN=(\alpha_sN_c/\pi)(dz/z)dq^2/q^2$ times the elementary $g^*+g\to \chi(1^+)$ cross section. That is
\be
\frac{d\hat\sigma}{dt}\Big|_{s\gg M^2,|t|}~=~\frac{dN}{dq^2}\hat\sigma(g^*+g\to \chi_c(1^+))\ .
\ee
(The factor $dz/z$ in $dN$ corresponds to the integration over the rapidity separation $y$ and is omitted here.)
In this way we get
\be
\hat\sigma(g^*+g\to \chi(1^+))=\frac{4\pi^2\alpha^2_sR^{'2}}{M^3}
|t|\frac{4M^2-2t}{m^8_t}\ ,
\label{eq:eq4}
\ee
where the virtuality, $q^2$, of off-mass-shell gluon $g^*$ plays the role of $t=q^2$ in (\ref{eq:eq1}).

Since the elementary cross section (\ref{eq:eq4}) vanishes as $q^2\to 0$ we cannot consider the incoming gluon, with momentum $q$, as an on-mass-shell parton. The inclusive cross section should therefore be written in terms of the {\em unintegrated} gluon density which is defined in such a way that
\be
xg(x,\mu_F)=\int^{\mu^2_F} f_g(x,q^2,\mu_F)\frac{dq^2}{q^2}\ .
\label{eq:ug}
\ee
Thus we obtain
\be
\frac{d\sigma}{dY}=\int x_1g(x_1,\mu_F)\hat\sigma(g^*+g\to\chi(1^+))f_g(x_2,q^2,\mu_F) \frac{dq^2}{q^2}~+~(x_2\leftrightarrow x_1)\ ,
\label{eq:ugg}
\ee
with $x_{1,2}=e^{\pm Y}\sqrt{(M^2-q^2)/s}$.

Note that the $dq^2/q^2$ integral does not now have a   logarithmic structure since the `hard' cross section (\ref{eq:eq4}) contains a factor $t=q^2$. In other words expression (\ref{eq:ugg}) should be considered as a NLO contribution (in comparison with LO $\chi(0^+,2^+)$ meson production) at the same level as the cross section (\ref{eq:eq2}). Moreover, it should not be considered as a new contribution -- it is just a part of the whole
cross section (\ref{eq:eq2}).

 The value of contribution (\ref{eq:ugg}) shown in Table 1 
is obtained by calculating the unintegrated gluon density starting from the MSTW2008~\cite{MSTW08} PDFs based on the KMR/MRW last-step prescription~\cite{KMR,MRW}.   In this prescription we have used the LO splitting functions but kept the more exact NLO kinematics. As it was shown in~\cite{MRW} this provides an accuracy close to that given by the NLO prescription.  Note that here we have used the gluons unintegrated 
over the virtuality, $q^2$ and not over the transverse momentum squared,
 $q^2_t$ (see~\cite{MRW} for details).

The contributions (\ref{eq:eq2}) and (\ref{eq:ugg}) were evaulated using the MSTW2008 LO PDFs  for the integrated gluons $xg$. For the very low $q^2<Q^2_0$ we assume the saturation-like behaviour
$xg(x,q^2)=xg(x,Q^2_0)(q^2/Q^2_0)$.

\section{Production via gluon reggeization \label{sec:Regge}}
\begin{figure}
\begin{center}
\includegraphics[height=6cm]{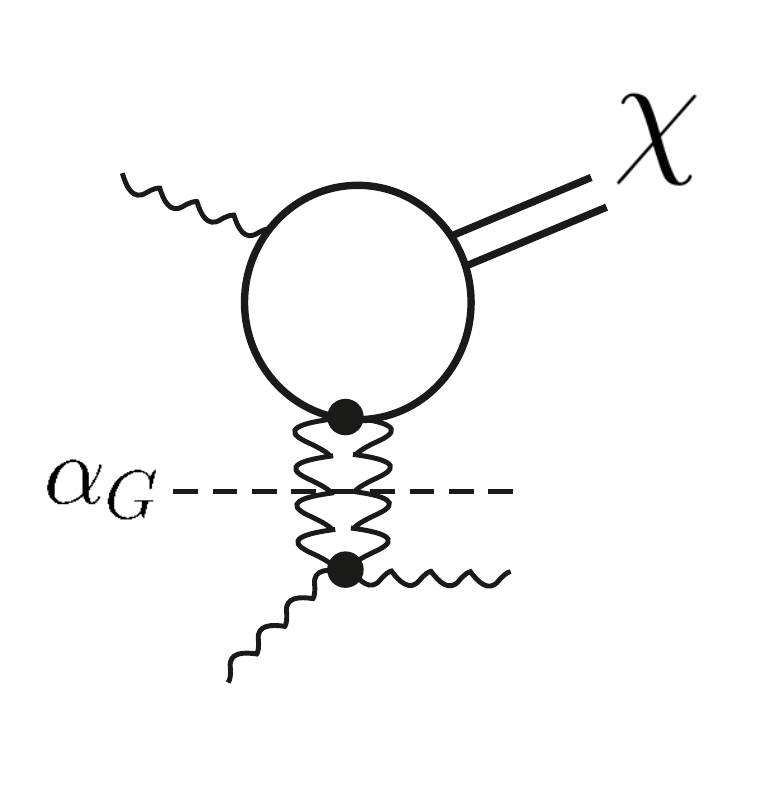}
\caption{\sf  $\chi(1^+)$ production via the fusion of an incoming gluon with a pair of gluons which form a Regge trajectory $\alpha_G$.}
\label{fig:regge}
\end{center}
\end{figure}
An attractive possibility to organize the fusion of one gluon with the {\em pair} of incoming  gluons, that is the $g+(gg)\to\chi(1^+)$ subprocess, is to consider the contribution coming from gluon reggeization, see Fig.~\ref{fig:regge}.  On one hand the gluon trajectory is described by the diagrams where the initial $t$-channel gluon is replaced by the exchange of two $t$-channel gluons. On the other hand this contribution is enhanced by large logarithms of the proton-proton energy, or to be more precise - by a $1/\omega_0$
 factor, where $\omega_0$ denotes the shift of the position of the BFKL vacuum 
 singularity $\alpha_{\rm BFKL}=1+\omega_0$. Recall that thanks to the 
bootstrap condition~\cite{BFKL} if we account for the gluon trajectory then we 
include {\em all} the contributions (of the antisymmmetric colour-octet states, which are of interest here) enhanced by the $1/\omega_0$ factor, that is - by the large leading logarithm
 of the energy.
 
 Let us explain in more detail how the gluon trajectory is built up. 
 Consider the diagrams where the additional $t$-channel gluon is added to the usual ladder diagram (inside the same parton cascade). The simplest such diagram (in Feynman gauge) is where the additional gluon is between the $\chi_c$ and the nearest s-channel gluon of the ladder (that is, between the $c$-quark loop and the gluon 3 in Fig.1b).
The contribution of this diagram is enhanced by two logarithms.
One logarithm, ln$q^2$, comes from the integration over the $k_t$ of the new $t$-channel gluon. The other logarithm comes from the integration over the longitudinal component of the new gluon's momentum, which corresponds to the integration over the mass of the intermediate $s$-channel gluon which emits the new $t$-channel gluon.
This longitudinal logarithm is actually equal to the rapidity separation
between the $\chi_c$ meson and the nearest $s$-channel gluon. This rapidity separation is driven by the intercept (the $x$-dependence) of the parton cascade. In the BFKL approach this logarithm is equal to $1/\omega_0$.
 Moreover, in terms of the BFKL amplitude, the coherent sum of such diagrams (where the new $t$-channel gluon couples to the different $s$-channel gluons in the ladder) is described by the gluon trajectory
 \be 
\alpha_G(q^2)=\frac{\alpha_sN_c}{2\pi}\ln q^2~.
\label{eq:g-tr}
\ee

Note that in this approach (just as in two gluon fusion $g^*+g \to\chi_c(1^+)$) the corresponding contribution vanishes when the `reggeized' gluon virtuality $q^2\to 0$. Indeed, due to gauge invariance, the whole set of diagrams which describe gluon reggeization can be reduced to a one-gluon loop inserted in the place of the original gluon propagator (see, for example,~\cite{GLR}).  Thus the result looks like cross section (\ref{eq:ugg}) multiplied by the gluon trajectory $\alpha_G(q^2)$
and by the $1/\omega_0$ factor.

Moreover, contrary to inclusive $J/\psi$ production ( where the gluon pair should be in a symmetric colour-octet state and the analogous contribution is imaginary), in the case of \ch1 we deal with the {\em antisymmetric} colour octet -- that is, with a true gluon trajectory of negative signature. 
Therefore the logarithmically enhanced contribution is real and {\em interferes} with the lowest $\alpha_s$ order $g^*+g\to\chi_c(1^+)$ amplitude considered in the previous section.

Thus, finally, to lowest order in $\alpha_s$, the cross section (\ref{eq:ugg}) should be multiplied by a `double logarithmic' factor $A$, where
\be
A=\left[1+2\frac{\alpha_sN_c}{2\pi\omega_0}\ln(q^2_1/q^2_2)\right]\ .
\label{eq:g-reg}
\ee
Here the $q^2$ logarithm comes from the integration over the transverse momentum $k_t$ in the loop corresponding to the gluon trajectory. In the region of $k_t^2\ll q^2$, the integral takes the form 
\be
\int^{q^2}\frac{dk^2_t}{k_t^2}\ .
\label{eq:ir-log}
\ee
 Note that the amplitude for unnatural parity $1^+$ meson production 
\be
g(q_1)+g(q_2)\to\chi(1^+)
\label{eq:permut}
\ee
is antisymmetric with respect to the permutation of the two gluons. Therefore in the amplitude (\ref{eq:permut}) 
we have the logarithm of the {\em ratio}
$q^2_1/q^2_2$. This result solves the problem of the infrared cutoff at low $k_t$ in the integral (\ref{eq:ir-log}). In the pure symmetric configuration (with $q_1^2=q^2_2)$ we get zero. This fact in some sense is similar to 
the Landau-Yang selection rule -- a spin $1^+$ particle cannot decay into two identical transverse gluons~\footnote{Recall that as we are looking for the leading logarithm the gluons may be considered as quasi-real, transverse particles.}. 

Recall that in the case of $\chi(1^+)$ production the second $t$-channel gluon (whose distribution is written here in terms of the `gluon reggeization' trajectory (\ref{eq:g-tr})) is not a virtual loop {\em correction} to the main amplitude. Rather, this is a particular new channel to produce a spin $1^+$ meson. For this reason it should not be combined with the real, $s$-channel
gluon emission. The infrared divergence in (\ref{eq:g-reg}) is cancelled between the diagrams with the upper ($q_1$) and the lower ($q_2$) gluon insertions;
and not between the virtual loop (reggeization) and the real gluon emission.\footnote{Just as in the case of the colourles Higgs boson production,
the infrared divergence caused by real emission is cancelled by the true reggeization diagram where (in Feynman gauge) the additional $t$-channel gluon couples to the $s$-channel gluons above and below the colourless boson.} 

The results of the numerical estimate of the corresponding order of $\alpha_s$ contribution\footnote{The contribution of the terms $\propto\ln q^2_1$ and $\propto\ln q^2_2$ are calculated separately. In the first case the gluon $q_1$ is written in terms of the unintegrated distribution, $f_g$, while for gluon $q_2$ we may use the integrated gluon distribution $x_2g(x_2,\mu_F)$, and vice versa.} are presented in the fourth column of Table 1.
Here we have used the value of $\omega_0=1/4$ which is close to that expected for the BFKL pomeron after the resummation of the next-to-leading log corrections~\cite{NLL}.  Recall that at the lowest $\alpha_s$ order, the `reggeized induced' contribution is proportional to $\alpha_s/\omega_0$. We choose reasonable values of $\omega_0$ and $\alpha_s$ to indicate the possible size of the effect.

As expected, the result is negative since at $q^2<0$ the gluon trajectory is shifted to lower values of $\alpha_G(q^2)<1$. However due to the cancellation between the $q_1$ and $q_2$ terms the whole contribution is not too large
 inspite of the $1/\omega_0$ enhancement.

 Recall however that at the present stage we account for the order of $\alpha_s$ reggeized correction only. When the "correction" becomes
 large the higher order $\alpha_s$ terms become important replacing effectively the first and negative $\alpha_s$ contribution (let denote it as $-\delta_R$) by the positive exponential factor like $\exp(-\delta_R)$.

\section{Production via two parton cascades}
\begin{figure}
\begin{center}
\includegraphics[height=6cm]{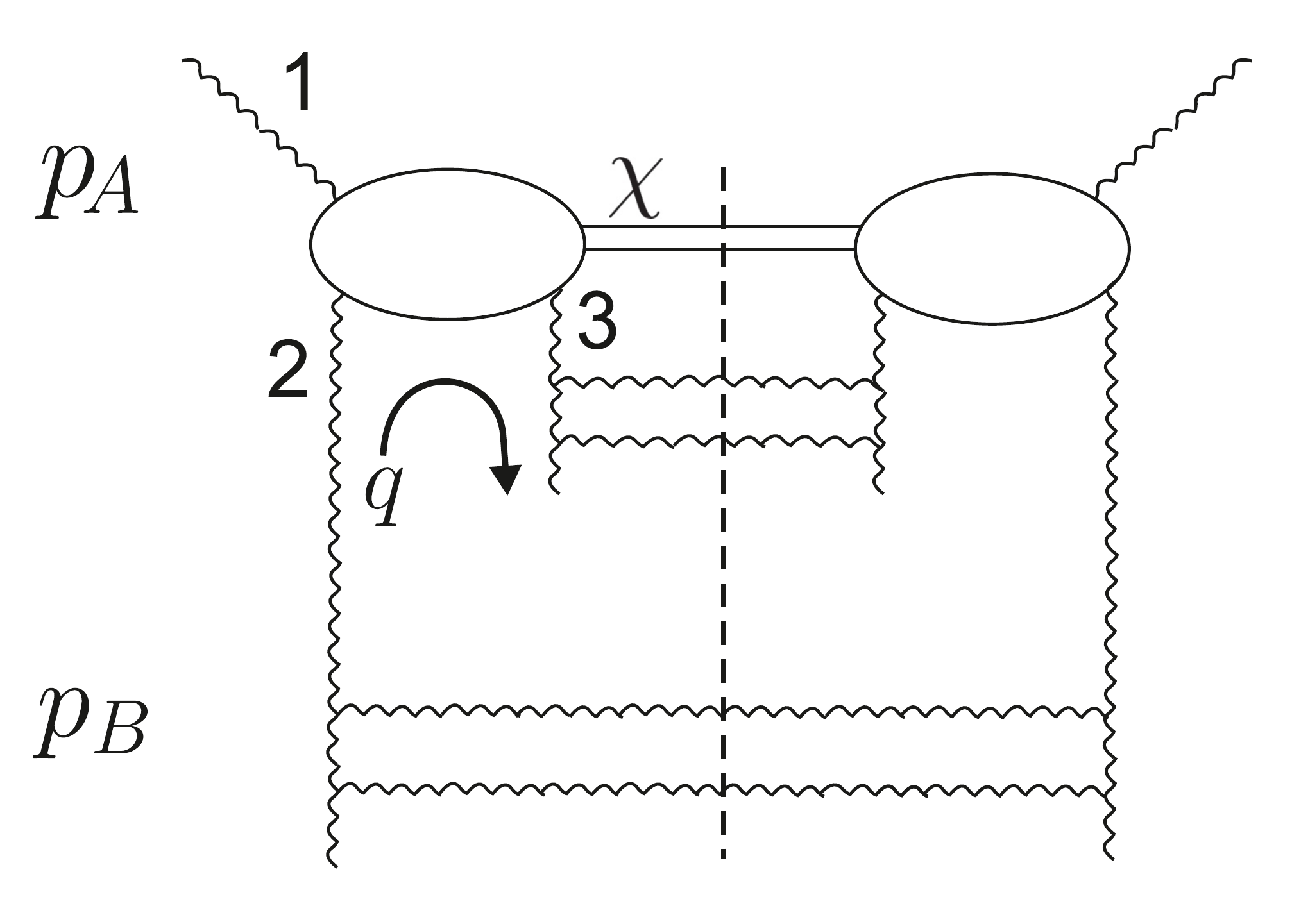}
\caption{\sf The diagram for the cross section for $\chi(1^+)$ production via the fusion of two gluons from different parton cascades, where $p_A$ and $p_B$ are the four momenta of the incoming protons (not shown).  The diagram is for the cross section, $AA^*$,  so the particles intersected by the dashed line are on-mass-shell.}
\label{fig:DPS}
\end{center}
\end{figure}
Here we consider the situation when all three (quasi-real) gluons couple to the heavy quark loop directly.
The most interesting possibility is to form the incoming $(gg)$ pair taking the two gluons from two different parton cascades, see Fig.~{\ref{fig:DPS}. The probability to find the corresponding pair is given by the product of the gluon densities,
$x_1g(x_1)\times x_2g(x_2)$, multiplied by the probability that the two cascades overlap in transverse space. Since in the low $x$ region (which is relevant at high energies) the parton density grows as a power of $(1/x)$ -- that is $xg(x)\propto x^{-\lambda}$ -- this contribution will dominate asymptotically at very high energies. 

In Double Parton Scattering (DPS) the probability of cascade overlap in transverse space is given by a factor $1/\sigma^{\rm  DPS}_{\rm eff}$ (see e.g.~\cite{FS}), where
\be
\frac 1{\sigma_{\rm eff}^{\rm DPS}}=\int F^4(q^2)\frac{d^2q}{4\pi^2}\ .
\label{eq:s-eff}
\ee 
Actually this factor results from the integration (\ref{eq:s-eff}) over the (`pomeron') loop formed by the two gluon cascades. The integral is driven by the proton form factor $F(t)$, that is by the $t\simeq q^2$ dependence of the proton-pomeron vertex. However, while for DPS the loop contains four form factors (leading to $F^4$ in (\ref{eq:s-eff})) in our case we have only two form factors (we need two cascades on one side of Fig.~{\ref{fig:DPS} only).
Therefore (assuming exponential dependence) we have 
\be
\sigma^\chi_{\rm eff}=\frac 12\sigma^{\rm DPS}_{\rm eff}\ .
\label{eq:s-chi}
\ee
 We  emphasize that the typical value of $|t|$ in the 
pomeron loop integration simultaneously plays the role of the lower limit for the  factorization scale, $\mu^2_F$. In terms of $\sigma^{\rm DPS}_{\rm eff}$ (and assuming the exponential $t$-behaviour) this
limit is  $\mu^2_F\ >\ \langle|t|\rangle=4\pi/\sigma^{\rm DPS}_{\rm eff}$.

Let us denote the momenta of the three gluons as $k_i$ ($i=1,2,3$).
In terms of the incoming proton's momenta, $p_A$ and $p_B$ we may write
\bea
k_1 &\simeq & \alpha p_A+k_{1t} \\
k_{i} & \simeq & \beta_{i} p_B +k_{it}\;\;\;\mbox{for}\;\; i=2,3.
\eea
To calculate the corresponding matrix element we use the gauge invariance condition (${\cal M}_\mu k_\mu=0 $) and replace the proton's momenta $p_A,p_B$ transferred through the spin part (or numerator) of gluon's propagators
by $k_{1t}/\alpha$ and $k_{it}/\beta_i$ respectively. Note that in order to keep the leading (DGLAP) logarithms in $k_t$ integrals we retain the lowest power of $k_{it}$ in the matrix element~\footnote{Together with the denominators ($1/k^2_{i}$)  of the gluon propagators this will give in the {\em  cross section} the logarithm $\int k^2_{it}d^2k_{it}/k^4_i$.}.

Thus the matrix element for $ggg\to \chi(1^+)$, corresponding to the heavy quark loop in Fig.~{\ref{fig:DPS},  takes a rather simple form
\be
{\cal M}=-iB f^{abc}\frac{A_sg^3}{N_c M} , 
\label{eq:m1}
\ee
where $f^{abc}$ is the antisymmetric colour tensor, $N_c=3$ is the number of colours, and $g$ is the QCD coupling ($\alpha_s=g^2/4\pi$). The basic amplitude $A_s$ contains a spin part (given by the trace around the quark loop) and a part corresponding to the propagator poles. It is of the form
\be
A_s=16(a_1z_1+a_2z_2)
\ee
where the trace gives
\bea
a_1 & = & (e_a\cdot k_{3t})~\epsilon_{\alpha\beta\gamma\delta}~k_{2t}^\alpha ~p_A^\beta ~p_B^\gamma ~e_\chi^\delta \\ 
a_2 & = & (e_a\cdot k_{2t})~\epsilon_{\alpha\beta\gamma\delta}~k_{3t}^\alpha ~p_A^\beta ~p_B^\gamma ~e_\chi^\delta\ ,
\eea
and the poles are
\bea
z_1 & = & \frac{2\alpha\beta_{23}~ s }{\alpha (2z-1)\beta_{23} s-|\vec k_{1t}-\vec k_{2t}|^2-M^2}  \\
z_2 & = & \frac{2\alpha \beta_{23}~ s }{\alpha\beta_{23} s(1-2z)-|\vec k_{1t}-\vec k_{2t}|^2-M^2},
\eea
where the denominators include the contribution of the longitudinal ($\alpha \beta_{23}s(1-2z)$) and transverse ($-|\vec k_{1t}-\vec k_{2t}|^2$) components of the square of the momentum.
Here $e_\chi$ and $e_a$ are the \ch1 and  $k_1$-gluon polarization vectors. Also we have introduced the relative momentum fraction $z=\beta_2/(\beta_2+\beta_3)$ where
\be
\beta_{23}=\beta_2+\beta_3=(
M^2+|\vec k_{1t}+\vec k_{2t}|^2)/\alpha s .
\ee
The value of $\beta_{23}$ is fixed by the \ch1 meson mass and its rapidity. Finally, the normalization constant $B$ in (\ref{eq:m1}) is given by
\be
B=\sqrt{\frac{3R^{'2}}{\pi M^3}}\ .
\ee

Recall that in the diagram for the cross section, Fig.~{\ref{fig:DPS}, we have four $t$-channel gluons from the proton $p_B$ side~\footnote{In general, there may be a diagram where the gluon pair $(gg)$ goes in the $p_B$ direction in one amplitude, $A$, but  goes in the other, $p_A$, direction in the amplitude $A^*$. Such a contribution {\it either} corresponds to the three gluon singularity (in the $\omega$ plane) which has a lower intercept ($\omega_0\sim 0$ or even negative -- this contribution is small): {\it or}, dealing with the BFKL pomeron, we have to consider this $(gg)$ pair as gluon reggeization which (in the lowest $\alpha_s$ order) was already considered in Sect.3.}. These 4 gluons form three loops of integration.
The integrals over the transverse momenta components have already discussed:
two loops gives the logarithmic integrals $\int k^2_{it}d^2k_{it}/k^4_i$ corresponding to the gluon PDF given by the parton cascade $i$, while the third integral (over $t$) is limited by the proton form factor. The longitudinal momentum component in the central loop is fixed by the condition that the \ch1 meson (and other secondary partons) should be on-mass-shell. Finally, we have the loop integrations over the gluon longitudinal momentum in each amplitude, $A$ and $A^*$. These integrals can be written in terms of $z$ and 
can be closed on the quark pole~\footnote{In general, in our case with a $L=1$ wave function, some part of the contribution may contain poles of second-order (see e.g.~\cite{kuhn}). However, as far as we keep just the leading logarithms in $k_t$, that is the lowest power of $k_t$, the residue of these second-order poles vanishes at $z=0$ or $z=1$. Therefore actually we deal only with simple first-order poles.}; that is, in our non-relativistic approximation (for the \ch1 wave function) we get an intermediate state with both quarks on-mass-shell. The pole position is 
\be
z=\frac{2(\vec{k}_{2t}\cdot \vec{k}_{3t})}{M^2}\ .
\label{eq:pole}
\ee 
So, finally, the matrix element (\ref{eq:m1}) reads
\be
{\cal M}=32\pi B f^{abc}\frac{g^3(a_1+a_2)}{N_c M}\ . 
\ee

To obtain the cross section we have to sum over the \ch1 polarizations, $e_\chi$, average over $e_a$ and integrate over the gluon transverse momenta. A problem is that according to (\ref{eq:pole}) the value of $z$ depends on these momenta and we cannot use the usual, integrated PDFs; each value of $k_{it}$ corresponds to its own $z$. Therefore we have to use the unintegrated distribution in the same way as described in Sect.~2.2.
Moreover in our calculation we have to keep only the poles with the positive $z<1$. Otherwise one of the gluons ($z$ or $1-z$) will have {\em negative energy} and so must be considered as an {\em outgoing} and not an incoming particle. Such a configuration describes the $gg\to \chi(1^+)+g$ subprocess which has already been discussed in Sect.~2.\\

Thus the `DPS' component of inclusive \ch1 cross section (which we call  $\sigma^{\rm DPS}$) reads
$$
\frac{d\sigma^{\rm DPS}}{dY}=\frac{\pi^3\alpha_s^3 R^{'2}}{\sigma^{\rm DPS}_{\rm eff}M^9}\frac{32}9
\int^{M^2}\frac{dk^2_{2t}}{k^2_{2t}}\frac{dk^2_{3t}}{k^2_{3t}}\alpha g(\alpha,M^2)$$
\be
\cdot\int^{2\pi}_0 \frac{d\phi}{2\pi} (1+\cos^2 \phi) f_g(x_2=\beta_{23} (1-z),k^2_{2t})
f_g(x_3=z\beta_{23},k^2_{3t})\Theta(z)\Theta(1-z)+(\alpha\leftrightarrow \beta_{23})\ ,
\ee
where $z$ is given by (\ref{eq:pole}) and where we take the factorization scale $\mu_F=M$. Since, within the Leading Logarithm approximation, both $k_{1t}^2,k^2_{2t}\ll M^2$, we may replace in the first argument of $f_g(\beta_{23}(1-z),...)$ the value of $(1-z)$ by 1, that is $f_g(\beta_{23}(1-z),...)$ becomes $f_g(\beta_{23},...)$. The origin of cos$^2 \phi$ in the cross section comes from the convolution of $a_1$ and $a_2$.
  
The predicted `DPS' part of the \ch1 production cross section is presented in the fifth column of Table 1. In the numerical calculation we have used MSTW2008LO partons~\cite{MSTW08} and $\sigma^{\rm DPS}_{\rm eff}=10$ mb~\cite{S-eff}.

\section{Discussion}
As  seen from Tables \ref{tab:1} and \ref{tab:2}, in the central rapidity region ($Y=0$) the main  contribution to \ch1 production comes from the most trivial $gg\to\chi(1^+)+g$ subprocess, even at 13 TeV. This contribution decreases with $|Y|$, that is, towards the edges of the available rapidity space. 

The expected correction due to gluon reggeization vanishes at $Y=0$, in accord with the generalized Landau-Yang selection rule, which is valid for the symmetric configuration. Moreover, for $Y\neq 0$ it is negative, and  at large rapidity $Y$ it becomes quite large in comparison with the original (without  reggeization) contribution. The absolute value of the Regge contribution increases with $|Y|$ (up to $Y\sim 5$)  making the overall rapidity ($Y$) distribution a bit narrower. We emphasize, however that the results of Tables \ref{tab:1} and \ref{tab:2} are presented just for illustration purposes, and include the ${\cal O}(\alpha_s)$ reggeized correction {\it only}. When the `correction' increases, the higher-order $\alpha_s$ terms become important replacing effectively the first and negative $\alpha_s$ contribution (which we denote as $-\delta_R$) by a positive exponential factor like $\exp(-\delta_R)$.

The DPS contribution, which originates from the fusion of gluons from two parton cascades, reveals quite a different behaviour. Due to the growth of the low $x$ gluon density
this part of cross section increases with the initial energy and/or with $|Y|$. The DPS mechanism for $\chi_c (1^+)$ production is seen to provide about $30\%$ of the cross section for $Y=4$ to 5.  At very high energies (asymptotically) 
 this contribution starts to dominate.  
 
 The values of the $\chi_b (1^+)$ cross sections are much smaller, but due to the larger $\chi_b$ mass the perturbative predictions are better justified. Note also the strong correction caused by gluon reggeization, which arises since the virtuality of the gluon is larger.

 \begin{table} [t]
\begin{center}
\begin{tabular}{|l||c|c|c||c|}\hline
 $Y$ &   $gg\to \chi +g$ & $g^*g\to \chi$ &  $G$-regge & DPS \\ \hline
        0.0   &     9.8 (17.2)   &      6.3 (10.5)    &     0.0 ~(0.0)  &      1.1 (2.2)\\
        1.0    &     9.5 (16.8)   &      6.2 (10.3)  &     -1.0 (-1.3)   &     1.2 (2.4)\\
        2.0   &     8.7 (15.6)  &     5.6 ~(9.6)   &    -1.9 (-2.5)   &    1.4 (2.9)\\
        3.0   &     7.4 (13.7)   &     4.7 ~(8.5)  &     -2.5 (-3.3)   &    1.7 (3.6)\\
        4.0   &     5.6 (11.3) &      3.6 ~(7.1)   &    -2.9 (-3.8)   &    2.1 (4.7)\\
        5.0   &     3.6  ~(8.5) &      2.1 ~(5.4)   &    -2.9 (-4.1)  &     2.3 (5.9)\\
        6.0   &     1.5 ~(5.1)  &      0.6 ~(2.9)  &     -2.4 (-4.5)    &    1.8 (4.6)\\
 \hline
\end{tabular}
\end{center}
\caption{\sf The cross section $d\sigma/dY$ in $\mu$b for producing $\chi_c(1^+)$ by the various mechanisms at 7 (13) TeV.  Note that the $g^*g\to \chi$ is already included in the   $gg\to \chi +g$ contribution, and has only been considered separately to facilitate the study of the Regge contribution. }
\label{tab:1}
\end{table}
 
  \begin{table} [t]
\begin{center}
\begin{tabular}{|l||c|c|c||c|}\hline
 $Y$ &   $gg\to \chi +g$ & $g^*g\to \chi$ &  $G$-regge & DPS \\ \hline
     0.0  &     37   &     30   &    ~ 0   &     2.2\\
    
    1.0   &     36    &     29   &    -10   &     2.4\\
    2.0   &     31    &     25   &    -18    &    2.7 \\
    3.0   &     24   &     18   &    -22   &     3.1\\
    
    4.0   &     16   &     11   &    -23   &     3.1\\
    
    5.0   &   ~8    &    ~3    &   -18    &    2.4\\
 \hline
\end{tabular}
\end{center}
\caption{\sf The cross section $d\sigma/dY$ in nb for producing $\chi_b(1^+)$ by the various mechanisms at 13 TeV. Note that the $g^*g\to \chi$ is already included in the   $gg\to \chi +g$ contribution, and has only been considered separately to facilitate the study of the Regge contribution. }
\label{tab:2}
\end{table}

\section*{Acknowledgments}
We are grateful to  Jean-Philippe Lansberg for a
useful discussion and valuable comments.
MGR thanks the IPPP at the University of Durham for hospitality. This work 
was supported by the RSCF grant 14-22-00281.
VAK thanks the Leverhulme Trust for an Emeritus Fellowship.

L.~N.~Lipatov and V.~S.~Fadin,
Sov. Phys. JETP {\bf 44}, 443 (1976);

E.~A.~Kuraev, L.~N.~Lipatov and V.~S.~Fadin,
Sov. Phys. JETP {\bf 45}, 199 (1977).

 
\bibitem{GLR} L.V. Gribov, E.M. Levin and M.G. Ryskin, Phys. Rept. {\bf 100} (1983) 1.

\bibitem{NLL} 
M.~Ciafaloni, D.~Colferai and G.P.~Salam, Phys. Rev. {\bf D60} (1999) 114036; \\ 
G.P. Salam, JHEP {\bf 9807}, 019 (1998);\\
V.A.~Khoze, A.D.~Martin, M.G.~Ryskin and W.J.~Stirling, Phys. Rev. {\bf D70} (2004) 074013.

\bibitem{FS} L. Ametller et al., Phys. Lett. {\bf B169} (1986) 289;\\
L.~Frankfurt and  M.~Strikman, Phys. Rev. {\bf D66} (2002) 031502.

\bibitem{kuhn}  
  J.H. Kuhn, J. Kaplan, El G.O. Safani, Nucl. Phys. 
{\bf B157} (1979) 125. 

\bibitem{S-eff}  
 CDF Collaboration (F. Abe  {\it et al.}) Phys. Rev. {\bf D56} (1997) 3811-3832,\\
  D0 Collaboration (V.M. Abazov  {\it et al.}),  Phys. Rev. {\bf D81} 
(2010) 052012.

\end{document}